\documentclass[11pt]{article}

\usepackage{amssymb,latexsym}
\usepackage{epsfig}
\usepackage{eufrak}
\usepackage{amsmath}
\usepackage{mathrsfs}
\usepackage{color}

\newtheorem{theorem}{Theorem}
\newtheorem{lemma}{Lemma}

\newtheorem{remark}{Remark}

\newcommand{\be}{\begin{equation}}
\newcommand{\ee}{\end{equation}}
\newcommand{\bee}{\begin{eqnarray*}}
\newcommand{\eee}{\end{eqnarray*}}
\newcommand{\bel}{\begin{eqnarray}}
\newcommand{\eel}{\end{eqnarray}}
\newcommand{\bec}{\begin{cases}}
\newcommand{\eec}{\end{cases}}
\newcommand{\bem}{\begin{bmatrix}}
\newcommand{\eem}{\end{bmatrix}}

\newcommand{\bed}{\begin{description}}
\newcommand{\eed}{\end{description}}
\newcommand{\bei}{\begin{itemize}}
\newcommand{\eei}{\end{itemize}}
\newcommand{\ben}{\begin{enumerate}}
\newcommand{\een}{\end{enumerate}}

\newcommand{\beL}{\begin{lemma}}
\newcommand{\eeL}{\end{lemma}}
\newcommand{\beT}{\begin{theorem}}
\newcommand{\eeT}{\end{theorem}}

\newcommand{\bpf}{\begin{pf}}
\newcommand{\epf}{\end{pf}}

\newcommand{\pfbox}{\hfill\mbox{$\Box$}}

\newenvironment{pf}{\paragraph*{Proof{\rm.}}}{\pfbox\bigskip}

\begin{document}

\title{{\bf Estimating Traffic Parameters with Rigorous Error Control}
\thanks{The author is currently with Department of Electrical Engineering,
Louisiana State University at Baton Rouge, LA 70803, USA, and Department of Electrical Engineering, Southern University and A\&M College, Baton
Rouge, LA 70813, USA; Email: chenxinjia@gmail.com}}

\author{Xinjia Chen}

\date{February, 2008}

\maketitle

\begin{abstract}

To perform a queuing analysis or design in a communications context, we need to estimate the values of the input parameters, specifically the
mean of the arrival rate and service time.  In this paper, we propose an approach for estimating the arrival rate of Poisson processes and the
average service time for servers under the assumption that the service time is exponential. In particular, we derive sample size (i.e., the
number of i.i.d. observations) required to obtain an estimate satisfying a pre-specified relative accuracy with a given confidence level. A
remarkable feature of this approach is that no a priori information about the parameter is needed. In contrast to conventional methods such as,
standard error estimation and confidence interval construction, which only provides post-experimental evaluations of the estimate, this approach
allows experimenters to rigorously control the error of estimation.

\end{abstract}

\section{Traffic Model}

The assumption of Poisson arrivals is usually valid in modeling traffic in communications \cite{stalling}.
There is well-developed mathematical theory justifying such an assumption.
Also, it is often assumed that the service time of a server in a queuing system is exponential.
In the case of telephone traffic, the service time is the time for
which a subscriber engages the equipment of interest.  In a packet-switching network,
the service time is the transmission time
and is therefore proportional to the packet length.  It is difficult to give
a sound theoretical reason why service times should be exponential,
but the fact in most cases they are very nearly exponential \cite{stalling}.
For design and analysis purpose, it is desirable to estimate accurately the load currently generated
by each device inter-connected in a network.
For example, the rate of packets generated by a terminal or the size of packets.

\section{Error Control}

Conventional advice recommends approximating the Poisson arrival rate $\lambda$ as $\frac{n}{T}$
where $n$ is the number of arrivals during an interval of time with length $T$.
The problem with this method is that
we do not know how accurate the estimate is.  Moreover, we have difficulty in choosing
the length $T$ to guarantee a certain accuracy.  Observing over-long time interval is wasteful.
On the other side, inadequate observation will lead to poor estimate.

Situations are similar for estimating
the average service time $\mu$.  The estimate of $\mu$ is calculated as
\[
\widehat{\mu} = \frac{ \sum_{i=1}^n X_i} {n},
\]
where $X_i, \; i=1, \cdots, n$ are i.i.d. observations of service time.  For the purpose of accuracy evaluation,
the standard deviation of $\widehat{\mu}$ is estimated as
\[
\widehat{\sigma} = \sqrt{ \frac{ \sum_{i=1}^n (X_i - \widehat{\mu})^2} {n-1} }.
\]
It should be noted that such an estimate of standard deviation is not a good measure of accuracy,
since itself is also a random variable.

An alternative of accuracy evaluation is confidence interval construction.  The limitation of
standard deviation estimation and confidence interval construction is that they are only post-experimental
evaluation of accuracy based on fixed number of observations.  These two conventional methods do not control
the accuracy of estimation because the error is determined once the number of experiments is fixed.
There are two natural methods of error control.
First, one can specify an absolute error bound $\varepsilon > 0$ and
confidence parameter $\delta \in (0,1)$ and determine the sample size $n$ such that
\[
\Pr \left\{ \left| \widehat{\eta} - \eta
 \right| < \varepsilon    \right\} > 1-\delta
\]
where $\widehat{\eta}$ is an estimate of the parameter $\eta$
(e.g., Poisson arrival rate or average service time).  Second,
one can specify a relative error bound $\epsilon \in (0,1)$
and confidence parameter $\delta \in (0,1)$ and determine the sample size $n$ such that
\[
\Pr \left\{  \left| \frac{ \widehat{\eta} - \eta}
{\eta}  \right| < \epsilon \right\} \geq 1 - \delta.
\]
The problem with the first method is that it is not feasible to determine the sample size
without a priori information of the parameter.  Moreover,
an absolute error bound is not a good indicator of the precision of the estimate.
The second method is better since a perfect measure would be expressed in terms of a relative error bound.
More importantly, the sample size can be determined without any information of the parameter to be estimated.

\section{Arrival Rate}
To estimate the arrival rate of a Poisson process, we have
\begin{theorem}  \label{Pora}
Define function $g(m,x):=e^{-x} \sum_{i=0}^{m-1} \frac{x^i}{i!}$.
Let $\epsilon, \; \delta  \in (0,1)$.  Let $n$ be the least integer such that
 \[
g \left( n, \frac{n}{1 + \epsilon} \right) - g \left( n, \frac{n}{1 - \epsilon} \right)
\geq 1-\delta.
\]
 Observe a sequence of $n$ interarrival times  $X_i, \;  i=1, \cdots, n$.
 Define $\widehat{\lambda} = \frac{ n} { \sum_{i=l}^n X_i}$. Then
\[
\Pr \left\{  \left| \frac{ \widehat{\lambda} - \lambda}
{\lambda}  \right| < \epsilon \right\} \geq 1 - \delta.
\]
\end{theorem}

\begin{pf}
Since all $n$ inter-arrival times are i.i.d. exponential random variables with parameter $\lambda$,
the characteristic function of the average inter-arrival time, denoted by
$\overline{X}= \frac{ \sum_{i=1}^n X_i} {n}$, is
$\phi_{\overline{X}}   (t) = ( 1- j \;\frac{t}{n \lambda})^{-n}$.
Let $Y= 2 n \lambda \overline{X} = \frac{2n \lambda } {\widehat{\lambda}} $.
Then $\phi_{Y} = (1- 2jt)^{-n}$
 which implies that $Y$ possesses a Chi-square distribution of degree $2n$, i.e., $\chi^2(2n)$.  Notice that
\[
\left| \frac{ \widehat{\lambda} - \lambda}
{\lambda}  \right| < \epsilon \;\;\; \Longleftrightarrow \;\;\;
\frac{ \widehat{\lambda}  } {  1 + \epsilon } <
\lambda < \frac{ \widehat{\lambda}   } { 1 - \epsilon } \;\;\; \Longleftrightarrow \;\;\; \frac{2n}{1 + \epsilon} <
Y < \frac{2n} {1 - \epsilon}.
\]
Thus
\[
\Pr \left\{  \left| \frac{ \widehat{\lambda} - \lambda}
{\lambda}  \right| < \epsilon \right\} = \Pr \left\{  \frac{2n}{1 + \epsilon} <
Y < \frac{2n} {1 - \epsilon} \right\}.
\]
Making use of the relation between the Chi-square distribution and the Poisson distribution (see \cite{Proakis}),
we have \[
\Pr \{ Y > x \} = g(n,x).
\]
Hence, by the definition of function $g(.,.)$,  we have
\[
\Pr \left\{  \frac{2n}{1 + \epsilon} <
Y < \frac{2n} {1 - \epsilon} \right\} =
g \left( n, \frac{n}{1 + \epsilon} \right) - g \left( n, \frac{n}{1 - \epsilon} \right) .
\]
The proof is thus completed.
\end{pf}

\begin{remark}
A remarkable fact is that no a priori information about the arrival rate $\lambda$
is required to compute the sample size for
pre-specified relative error bound and confidence level.
\end{remark}

\begin{figure}[htbp]
\centerline{\psfig{figure=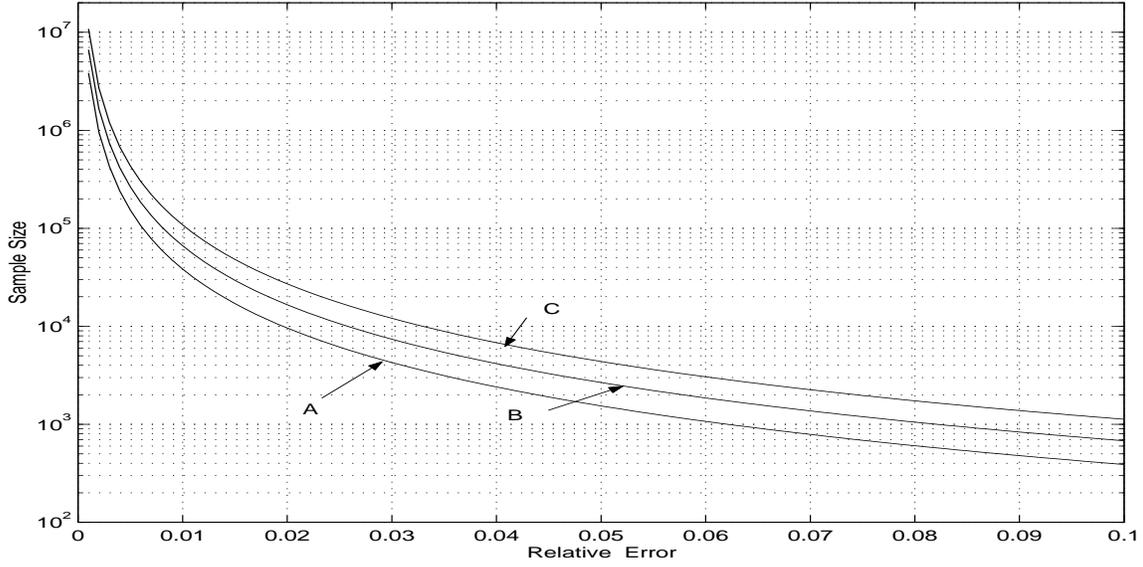, height=3in, width=6.0in
}} \caption{ Estimating Poisson Arrival Rate.   Plot A corresponds to $\delta = 0.05$,
plot B corresponds to $\delta = 0.01$, plot C corresponds to $\delta = 0.001$. }
\label{fig1}
\end{figure}

Figure~\ref{fig1} shows the sample size required to
obtain an estimate of the Poisson arrival rate with pre-specified relative accuracy (quantified by relative error bound $\epsilon$)
and confidence level.
It can be seen from Figure 1 that, to obtain an estimate $\widehat{\lambda}$ such that
\[
\Pr \left\{  \left| \frac{ \widehat{\lambda} - \lambda}
{\lambda}  \right| < 0.01 \right\} = 0.99,
\]
 we need to observe $n = 6.6 \times 10^4$ inter-arrival times
$X_i, \;  i=1, \cdots, n$ and compute $\widehat{\lambda} = \frac{ n} { \sum_{i=l}^n X_i}$.

\section{Average Service Time}
To estimate the average of exponential service time, we have
\begin{theorem}  \label{explam}
Let $\epsilon, \; \delta  \in (0,1)$.  Let $n$ be the least integer such that
 \[
g \left( n, \frac{n (1 + \epsilon)}{1 + 2 \epsilon} \right) - g \left( n, n(1 + \epsilon) \right)
\geq 1-\delta.
\]
Let $X_i,  \; i=1, \cdots, n$ be i.i.d. observations of service time.
 Define $\widehat{\mu} = \frac{ \sum_{i=l}^n X_i} {n}$. Then
\[
\Pr \left\{  \left| \frac{ \widehat{\mu} - \mu}
{\mu}  \right| < \epsilon \right\} > 1 - \delta.
\]
\end{theorem}

\begin{pf}
Let $Z= 2 n \lambda \overline{X} = \frac{2n \widehat{\mu}} {\mu} $.
Then $Z$ possesses a Chi-square distribution of degree $2n$, i.e., $\chi^2(2n)$.  Notice that
\[
\left| \frac{ \widehat{\mu} - \mu}
{\mu}  \right| < \epsilon \;\;\; \Longleftrightarrow \;\;\;
\frac{ \widehat{\mu}  } {  1 + \epsilon } <
\mu < \frac{ \widehat{\mu}   } { 1 - \epsilon }.
\]
Moreover,
\[
\frac{ \widehat{\mu}  } {  1 + \epsilon } <
\mu < \frac{ \widehat{\mu}   } { 1 - \epsilon } \;\;\;
\Longrightarrow \;\;\; \frac{2n(1 + \epsilon)}{1 +2 \epsilon} <
Z < 2n (1 + \epsilon)
\]
because
\[
\frac{1 + \epsilon}{1 +2 \epsilon} > 1 - \epsilon.
\]
It follows that
\[
\Pr \left\{  \left| \frac{ \widehat{\mu} - \mu}
{\mu}  \right| < \epsilon \right\} \geq \Pr \left\{ \frac{2n(1 + \epsilon)}{1 +2 \epsilon} <
Z < 2n (1 + \epsilon)
\right\}.
\]
Making use of the relation between the Chi-square distribution and the Poisson distribution,
we have $\Pr \{ Z > x \} = g(n,x)$.  Recall the definition of function $g(.,.)$, we have
\[
\Pr \left\{ \frac{2n(1 + \epsilon)}{1 +2 \epsilon} <
Z < 2n (1 + \epsilon)
\right\} = g \left( n, \frac{n (1 + \epsilon)}{1 + 2 \epsilon} \right) - g \left( n, n(1 + \epsilon) \right).
\]
The proof is thus completed.
\end{pf}

\begin{remark}
It can be seen that no a priori information about the average service time $\mu$
is required to compute the sample size for
pre-specified relative error bound and confidence level.
\end{remark}

\begin{figure}[htbp]
\centerline{\psfig{figure=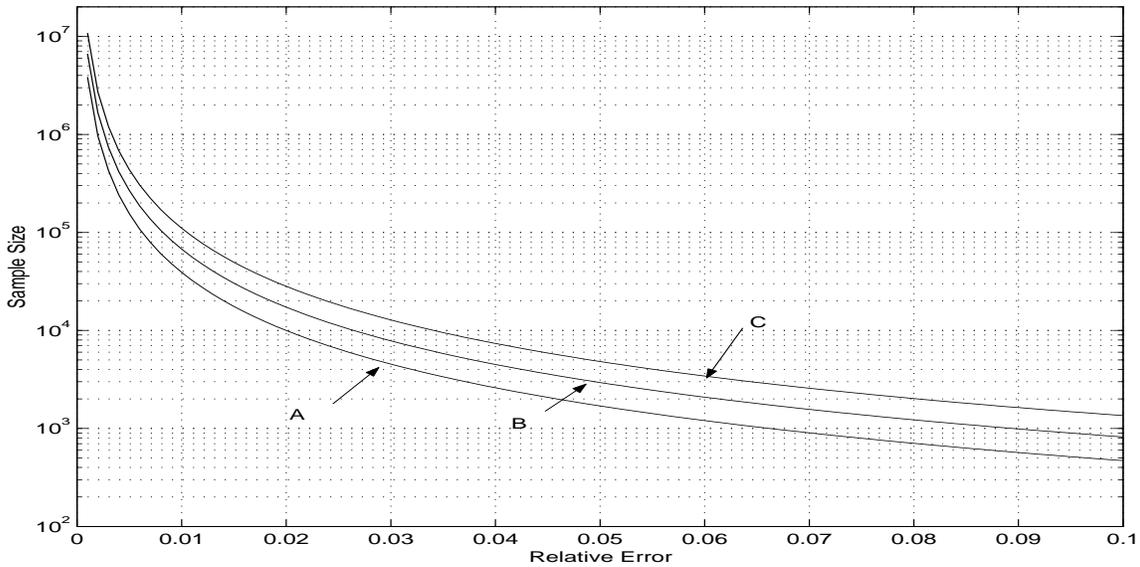, height=3in, width=6.0in
}} \caption{ Estimating Average Service Time.  Plot A corresponds to $\delta = 0.05$,
plot B corresponds to $\delta = 0.01$, plot C corresponds to $\delta = 0.001$. }
\label{fig2}
\end{figure}

Figure~\ref{fig2} shows the sample size required to
obtain an estimate of the average service time with pre-specified relative accuracy
(quantified by relative error bound $\epsilon$) and confidence level.
It can be seen from Figure 2 that, to come up with an estimate $\widehat{\mu}$ such that
\[
\Pr \left\{  \left| \frac{ \widehat{\mu} - \mu}
{\mu}  \right| < 0.05 \right\} > 0.95,
\]
we need to obtain the average of $n = 1.8 \times 10^3$ i.i.d. observations of service time.

\end{document}